\documentclass[a4paper,11pt]{article}
\usepackage{axodraw}

\usepackage{jheppub} 

\usepackage[T1]{fontenc} 
\title{\boldmath A four-dimensional approach to quantum field theories}

\author{R. Pittau}
\affiliation{Departamento de F\'{\i}sica Te\'orica y del Cosmos,\\
Campus Fuentenueva s. n., Universidad de Granada 
E-18071 Granada, Spain}
\emailAdd{pittau@ugr.es}

\abstract{I present a novel Four-Dimensional Regularization/Renormalization approach (FDR) to ultraviolet divergences in field theories which can be interpreted as a natural separation between physical and non physical degrees of freedom. Based on the observation that some infinities can be reabsorbed into the vacuum expectation value of the fields, rather than into the parameters of the Lagrangian, a new type of four-dimensional loop integral is introduced (the FDR integral) which is independent of any UV regulator and respects all properties required by gauge invariance. FDR reproduces the correct ABJ anomaly and no change in the definition of $\gamma_5$ is needed. With FDR the possibility is open for an approach to UV infinities in which the renormalization program is substituted by a simple reinterpretation of the appearing loop integrals as FDR ones, leading to important consequences in the context of non-renormalizable field theories. Finally, I show how FDR can also be used to regularize infrared and collinear divergences.}

\begin{document} 

\newcommand{\mur}{\mu_{\scriptscriptstyle R}}
\newcommand{\bqa}{\begin{eqnarray}}
\newcommand{\eqa}{\end{eqnarray}}

\maketitle
\flushbottom
\section{Introduction}
\label{sec:intro}
The study of the ultraviolet (UV) infinities occurring in quantum field theories has a long story~\cite{Dyson:1949ha,Dyson:1949bp,Salam:1951sm,Salam:1951sj}. The commonly accepted interpretation is that they occur in intermediate steps of loop calculations, but have no physical meaning, being reabsorbable into a redefinition of the parameters of the theory, which should therefore be readjusted order by order.
The proof that this can be carried out in a way consistent with the symmetries of the theory is at the base of the renormalization program, and requires to regularize the divergences by means of a regulator, showing that the dependence on the regulator drops out in physical observables.
A mathematical proof of the feasibility of this approach is due to the Bogoliubov,  Parasiuk~\cite{Bogoliubov:1957gp}, Hepp~\cite{Hepp:1966eg} and Zimmermann~\cite{Zimmermann:1969jj} (BPHZ). In the BPHZ renormalization scheme, the divergent Green functions are Taylor expanded up to the order needed to reach convergent integrals, and the renormalization conditions are imposed directly on the finite terms of the expansion. A key ingredient for the success of this procedure is showing that overlapping divergences, which cannot be compensated by local counterterms, cancel.
When applied to gauge theories, BPHZ may break gauge invariance in the intermediate steps, and is technically quite involved. Therefore, other alternatives, more manageable from a technical point of view, have been proposed, such as 
Pauli Villars~\cite{Pauli:1949zm}, or Speer regulators~\cite{Speer:1972wz}, which, however, may lead to gauge dependent results. This is not yet considered to be a serious drawback, since the missing pieces could be reinserted back by enforcing the Ward-Slavnov-Taylor
identities of the theory. The situation can then be summarized as follows: the symmetry of the theory is a guideline to prove the correctness of the result. 

In general, restoring gauge invariance may be technically difficult and error-prone, so when Dimensional Regularization (DR) emerged~\cite{'tHooft:1972fi}, which automatically respects gauge symmetries, the proof of the renormalizability of the Yang-Mills theories, with and without spontaneous symmetry breaking, could be carried out, and their predictive power started being exploited.
 However, the DR requirement of working at a dimensionality different from the physical one led to several attempts to find alternative four-dimensional solutions to the UV problem, such as differential renormalization~\cite{Freedman:1991tk}, constrained differential renormalization~\cite{delAguila:1997kw,delAguila:1998nd}, which both work in the coordinate space, and implicit renormalization~\cite{Battistel:1998sz,Cherchiglia:2010yd}, directly applicable in the momentum space.

A problem much more severe than the choice of the regularization procedure is represented by field theories which diverge so badly that new infinities are generated order by order in the perturbative expansion, which cannot be reabsorbed into the Lagrangian, at least with a finite number of couterterms. Such theories are called non-renormalizable, and are commonly interpreted as effective ones, the fundamental truth being still to be unrevealed.

In this paper, I present an interpretation of the UV infinities as unphysical degrees of freedom, which have to decouple. Elaborating on its consequences, I introduce an approach, dubbed Four-Dimensional Regularization/Renormalization (FDR), in which a key role is played by the definition of a four-dimensional, regulator free, gauge-invariance preserving integral, which allows one to directly compute renormalized Green Functions with no need of reabsorbing infinities into the Lagrangian. When applied to non-renormalizable theories, FDR leads to a possible interpretation in which they could acquire some predictivity.

The outline of the paper is as follows. Section~\ref{sec:sec1} explains how UV divergences and physical degrees of freedom can be disentangled. In sections~\ref{sec:fdr}, \ref{sec:1loop} and \ref{sec:2loop} the FDR integral is introduced, and the one-loop and two-loop cases discussed in detail.
In section~\ref{sec:abj} I show how the ABJ anomaly is naturally predicted by FDR, and comment on the role of $\gamma_5$.
The FDR renormalization program is discussed in section~\ref{sec:ren}, together 
with the aforementioned interpretation of the non-renormalizable theories.
Finally, the use of FDR to regularize infrared and collinear divergences is presented in section~\ref{sec:icol}.

\section{The FDR approach to UV divergences} 
\label{sec:sec1}
 A simple redefinition of the vacuum may reabsorbe some of the UV infinities occurring at large values of the integration momentum.
To illustrate this phenomenon, I start with the simplest possible example, namely a scalar theory with cubic interaction \footnote{To be considered part of a more complete theory.}
\bqa
{\cal L}= \frac{1}{2}(\partial_\mu \Phi)^2 -\frac{M^2}{2} \Phi^2
-M\frac{\lambda}{3!} \Phi^3\,. 
\eqa
The tadpole contribution generated at one-loop is usually compensated by introducing an {\em ad hoc} counterterm in the Lagrangian.
Instead of following this procedure, I perform a global field shift \footnote{Assuming no normal ordering for the Lagrangian.} 
\bqa
\label{eq:shift}
\Phi \to \Phi + v\,,
\eqa
which reproduces the usual Feynman rules plus extra ones, as drawn in figure~\ref{fig:fr3}.
\begin{figure}
\begin{center}
\begin{picture}(300,140)(0,0)

\SetOffset(35,100)
\Line(-60,0)(-20,0)
\LongArrow(-50,5)(-30,5)
\Text(-43,10)[l]{$ p $}
\Text(-15,0)[l]{$ \displaystyle~~ =~~ \frac{i}{p^2- M^2} $ ,}

\SetOffset(245,100)
\Line(-60,0)(-20,0)
\Line(-20,0)(15,20)
\Line(-20,0)(15,-20)
\Text(15,0)[l]{$ \displaystyle~~ =~ -~ i {\lambda}M $ ,}

\SetOffset(40,17)
\Line(-65,12)(-35,12)
\BCirc(-30,12){5}
\Line(-33,15)(-27,9)
\Line(-33,9)(-27,15)
\Text(-20,12)[l]{$ \displaystyle~~ =~ -~ i \left(M^2v + \frac {{\lambda}M}{2}v^2\right) $ ,}

\SetOffset(250,17)
\Line(-65,12)(-35,12)
\BCirc(-30,12){5}
\Line(-33,15)(-27,9)
\Line(-33,9)(-27,15)
\Line(-25,12)(0,12)
\Text(5,12)[l]{$ \displaystyle~~ =~ -~ i {\lambda}Mv $ .}

\end{picture}
\caption{\label{fig:fr3} Feynman rules for the $\lambda \Phi^3$ theory (top) and extra vertices generated by the shift $\Phi \to \Phi + v$ (bottom).}
\end{center}
\end{figure}
The only divergent diagrams at one-loop are the tadpole $T$ and the 2-point function 
$i\,\Sigma$
\bqa
T = \frac{\lambda M}{2}  I_2\,,~~~i\,\Sigma =  \frac{\lambda^2 M^2}{2} I_0\,.
\eqa
The one-loop scalar integrals \footnote{$\mur$ denotes the renormalization scale.}
\bqa
I_2 = \mur^{-\epsilon} \int d^n q\frac{1}{D}\,,~~~~~I_0 = \mur^{-\epsilon} \int d^n q\frac{1}{DD_p} \,,
\eqa
with
\bqa
D = (q^2-M^2)\,,~~~D_p = ((q+p)^2-M^2)\,,
\eqa
can be computed in $n= 4+\epsilon$ dimensions as
\bqa
\label{eq:eqlim02}
I_2 &=& \lim_{\mu \to 0} \mur^{-\epsilon} \int d^n q\frac{1}{\bar D}\,,   \nonumber \\ 
I_0 &=& \lim_{\mu \to 0} \mur^{-\epsilon} \int d^n q\frac{1}{\bar D\bar D_p}\,,
\eqa
where a small extra mass $\mu$ has been introduced in the propagators
\bqa
\label{eq:den}
\bar D   = D-\mu^2\,,~~~\bar D_p = D_p-\mu^2\,. 
\eqa
That allows one to use the partial fraction identities
\begin{eqnarray}
\label{eq:id}
\frac{1}{\bar D}  &=& \frac{1}{\bar{q}^2} \left(1 + \frac{M^2}{\bar D} \right)\,,\nonumber \\
\frac{1}{\bar D_p} &=& 
\frac{1}{\bar{q}^2} \left(1 + \frac{d(q)}{\bar D_p} \right)\,,
\end{eqnarray}
where
\begin{eqnarray}
\bar q^2 = q^2-\mu^2\,,~~~d(q)   = M^2-p^2-2 (q \cdot p) \,,
\end{eqnarray}
to rewrite the denominators as
\bqa
\label{eq:eqden1}
\frac{1}{\bar D} &=& \frac{1}{\bar q^2} 
                + \frac{M^2}{\bar q^4}
               + \frac{M^4}{\bar D \bar q^4}
\eqa
and
\bqa
\label{eq:eqden2}
\frac{1}{\bar D \bar D_p} &=& \frac{1}{\bar q^4} 
              + \frac{d(q)}{\bar q^4 \bar D_p}
              + \frac{M^2}{\bar q^2 \bar D \bar D_p}\,.
\eqa
In this way, the UV divergences are moved to the first two terms of
eq.~\ref{eq:eqden1} and to the first one of eq.~\ref{eq:eqden2} \footnote{Without $\mu$ the remaining terms would generate infrared divergences.}, and
since
\bqa
\lim_{\mu \to 0} ~\mur^{-\epsilon} \int d^n q\frac{1}{\bar q^2} = 0\,,
\eqa
only the logarithmically divergent integral
\bqa
\mur^{-\epsilon} \int d^n q\frac{1}{\bar q^4} \equiv i\, I_{inf}
\eqa
remains. The divergent parts are then computed as
\bqa
\label{eq:eqcanc}
\left. T        \right|_{inf} = i \lambda \frac{M^3}{2}  I_{inf}\,,~~~
\left. i \Sigma \right|_{inf} = i \lambda^2 \frac{M^2}{2}   I_{inf}\,,
\eqa 
in terms of a common integral which no longer depends on
any physical scale.
One can now define the $\Phi$ vacuum in such a way that no tadpoles occur,
as in figure~\ref{fig:notad}, which fixes the divergent part of $v$ to the value
\bqa
v_{inf} = \lambda M \frac{I_{inf}}{2} + {\cal O}(\lambda^3)\,.
\eqa

\begin{figure}
\begin{center}
\begin{picture}(300,40)(0,0)

\SetOffset(130,0)
\Line(-65,12)(-35,12)
\BCirc(-30,12){5}
\Line(-33,15)(-27,9)
\Line(-33,9)(-27,15)
\Text(-20,12)[l]{$ \displaystyle~~ +  $}
\SetOffset(200,0)
\Line(-65,12)(-35,12)
\BCirc(-25,12){10}
\Text(-10,12)[l]{$ \displaystyle~~ =~~ 0$}

\end{picture}
\caption{\label{fig:notad}
The no-tadpole condition on the shifted $\lambda \Phi^3$ scalar theory.
}
\end{center}
\end{figure}
When including the extra vertex in the computation of the $\Phi$ self energy
$\bar \Sigma$, as in figure~\ref{fig:sigma3}, this is just what one needs to cancel its UV behavior
\bqa
\left. i \bar \Sigma \right|_{inf} = 0\,.
\eqa
\begin{figure}
\begin{center}
\begin{picture}(300,50)(0,0)

\SetOffset(110,0)
\Text(-85,12)[l]{$ \displaystyle~~ i {\bar \Sigma}~~= $}
\Line(-35,12)(-5,12)
\BCirc(5,12){10}
\Line(15,12)(45,12)
\Text(50,12)[l]{$ \displaystyle~~ +~~$}
\SetOffset(270,0)
\Line(-65,12)(-35,12)
\BCirc(-30,12){5}
\Line(-33,15)(-27,9)
\Line(-33,9)(-27,15)
\Line(-25,12)(0,12)

\end{picture}
\caption{\label{fig:sigma3} The complete $\Phi$ self-energy in
the shifted $\lambda \Phi^3$ theory.}
\end{center}
\end{figure}
Notice that both divergences have been removed by fixing just one parameter \footnote{In agreement with ref.~\cite{Soln:1988rc}.}, while the traditional approach would have required two counterterms. As will be explained later, the dependence on $\mu$ obtained by integrating in four dimensions the left over terms in eqs.~\ref{eq:eqden1} and~\ref{eq:eqden2} can only be logarithmic, and can be traded for the usual dependence on $\mur$.

In a scalar theory with quartic interaction 
\bqa
{\cal L}= \frac{1}{2}(\partial_\mu \Phi)^2 -\frac{M^2}{2} \Phi^2
-\frac{\lambda}{4!} \Phi^4\,, 
\eqa
no tadpole is generated at one-loop. Nevertheless, one can still reabsorbe the mass renormalization into the vacuum expectation value of $\Phi$. The same shift of eq.~\ref{eq:shift} generates now the extra Feynman rules given in figure~\ref{fig:fr4}. 
\begin{figure}
\begin{center}
\begin{picture}(300,150)(0,0)

\SetOffset(50,100)
\ArrowLine(-60,0)(-20,0)
\LongArrow(-50,5)(-30,5)
\Text(-43,10)[l]{$ p $}
\Text(-15,0)[l]{$ \displaystyle~~ =~~ \frac{i}{p^2- M^2} $ ,}   

\SetOffset(250,100)
\Line(-30,15)(-10,-15)
\Line(-30,-15)(-10,15)
\Text(-10,0)[l]{$ \displaystyle~~ =~-~i{\lambda} $ ,}

\SetOffset(0,20)
\BCirc(-30,12){5}
\Line(-33,15)(-27,9)
\Line(-33,9)(-27,15)
\Line(-30,7)(-30,-12)
\Line(-54,25)(-35,15)
\Line(-25,15)(-6,25)
\Text(-10,12)[l]{$ \displaystyle = - i{\lambda}v $ ,} 

\SetOffset(115,20)
\Line(-60,12)(-35,12)
\BCirc(-30,12){5}
\Line(-33,15)(-27,9)
\Line(-33,9)(-27,15)
\Line(-25,12)(0,12)
\Text(7,13.5)[l]{$ \displaystyle = -i \frac{{\lambda}v^2}{2} $ ,}

\SetOffset(260,20)
\Line(-60,12)(-35,12)
\BCirc(-30,12){5}
\Line(-33,15)(-27,9)
\Line(-33,9)(-27,15)
\Text(-20,12)[l]{$ \displaystyle = - i \left(M^2v + \frac {{\lambda}}{6}\,v^3\right) $ .}
\end{picture}
\caption{\label{fig:fr4}
Feynman rules for the $\lambda \Phi^4$ theory (top) and extra vertices generated by the shift $\Phi \to \Phi + v$ (bottom).}
\end{center}
\end{figure}
\begin{figure}
\begin{center}
\begin{picture}(300,60)(0,0)

\SetOffset(50,15)
\GBox(-34,11)(-28,18){0}
\Line(-31,11)(-31,-12)
\Line(-54,26)(-34,16)
\Line(-28,16)(-8,26)
\Text(0,12)[l]{$ \displaystyle~~ =~ $} 
 
\SetOffset(150,15)
\BCirc(-30,12){5}
\Line(-33,15)(-27,9)
\Line(-33,9)(-27,15)
\Line(-30,7)(-30,-12)
\Line(-54,25)(-35,15)
\Line(-25,15)(-6,25)
\Text(0,12)[l]{$ \displaystyle~~ +~ $}

\SetOffset(240,15)
\Line(-30,12)(-30,-12)
\Line(-54,25)(-30,12)
\Line(-30,12)(-6,25)

\SetOffset(260,3)
\BCirc(-30,12){5}
\Line(-33,15)(-27,9)
\Line(-33,9)(-27,15)
\Line(-54,26)(-35,15)
\Text(0,25)[l]{$ \displaystyle~~ =~\frac{i{\lambda^2}}{6}~\frac{v^3}{M^2} $} 
\end{picture}
\caption{\label{fig:ident} A useful identity for 
the shifted $\lambda \Phi^4$ theory.}
\end{center}
\end{figure}
Due to the identity of figure~\ref{fig:ident}, 3-point vertices and tadpoles do not occur at ${\cal O}(\lambda)$, thus the $\Phi$ self-energy is simply obtained by summing the one-loop diagram and the
2-point extra vertex, as in figure~\ref{fig:sigma4}, giving
\bqa
\left. i \bar \Sigma \right|_{inf} = i\frac{\lambda}{2} (M^2 I_{inf} -v_{inf}^2)\,.
\eqa
The value $v_{inf}^2= M^2 I_{inf}$ is such that no infinite mass renormalization is needed. This solution is acceptable perturbatively, but it does not represent a correction ${\cal O}(\lambda)$ around $<\phi> = 0$, as in the previous case.
One can impose vacuum stability at higher orders by spontaneously breaking the symmetry with the condition
\noindent 
\bqa
M^2+\frac{\lambda v^2}{6} = 0\,,
\eqa
and repeating the exercise with this new vacuum. 

\begin{figure}
\begin{center}
\begin{picture}(300,50)(0,0)

\SetOffset(120,0)
\Text(-85,12)[l]{$ \displaystyle~~ i {\bar \Sigma}~~= $}
\Line(-30,10)(20,10)
\GOval(-5,23)(13,7)(0){1}
\Text(38,10)[l]{$ \displaystyle + $}
\SetOffset(250,-2)
\Line(-65,12)(-35,12)
\BCirc(-30,12){5}
\Line(-33,15)(-27,9)
\Line(-33,9)(-27,15)
\Line(-25,12)(0,12)

\end{picture}
\caption{\label{fig:sigma4}
The complete $\Phi$ self-energy in
the shifted $\lambda \Phi^4$ scalar theory.}
\end{center}
\end{figure}

\begin{figure}
\begin{center}
\begin{picture}(300,170)(0,0)

\SetOffset(80,87)
\BCirc(-30,12){20}
\Line(-30,32)(-30,12)
\Line(-30,12)(-15,-1)
\Line(-30,12)(-45,-1)
\DashLine(-50,10)(-75,0){2}
\Photon(-30,-8)(-30,-33){2}{6}
\Gluon(-20,29)(5,48){2}{4}
\Gluon(-65,48)(-40,29){2}{4}
\Line(-30,22)(-13,14)
\Line(-8,12)(12,3)
\Gluon(-50,40)(-30,56){2}{4}
\Text(-23,-45)[tr]{(a)}

\SetOffset(220,139)
\SetWidth{1.55}
\BCirc(-30,12){20}
\Line(-30,32)(-30,12)
\Line(-30,12)(-15,-1)
\Line(-30,12)(-45,-1)
\Text(-23,-20)[tr]{(b)}

\SetOffset(220,35)
\SetWidth{0.5}
\CArc(-30,12)(20,-40,220)
\Line(-30,32)(-30,12)
\Line(-30,12)(-15,-1)
\Line(-30,12)(-45,-1)
\DashLine(-50,10)(-75,0){2}
\Gluon(-20,29)(5,48){2}{4}
\Gluon(-65,48)(-40,29){2}{4}
\Line(-30,22)(-13,14)
\Line(-8,12)(12,3)
\Gluon(-50,40)(-30,56){2}{4}
\Text(-23,-25)[tr]{(c)}

\SetWidth{1.55}
\Line(-30,5)(-15,-8)
\Line(-30,5)(-45,-8)
\CArc(-30,5)(20,220,320)
\end{picture}
\caption{\label{fig:vacuum} Generic diagram contributing to the interaction (a). {\em Vacuum diagrams} generated when all integration momenta are large (b)
and when one sub-loop integration momentum goes to infinity (c).}
\end{center}
\end{figure}
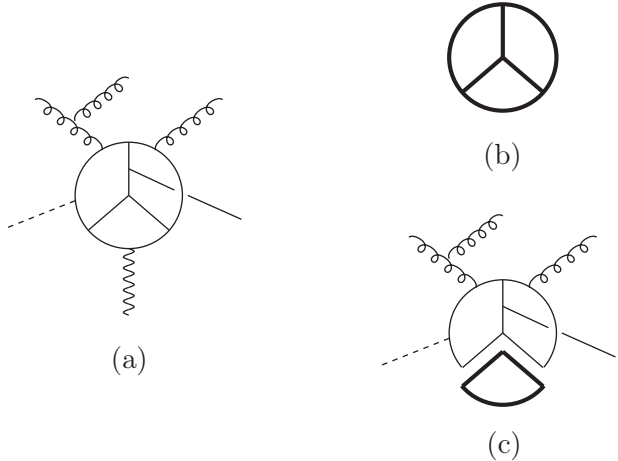

The lesson to be learnt from these examples is that unphysical degrees of freedom occurring in UV divergent loop integrals decouple once observable physics is described relative to the vacuum. 
 In addition, a parametrization as in
eqs.~\ref{eq:eqden1} and~\ref{eq:eqden2}, in which the UV part is moved to an integral $I_{inf}$, which does not depend on any physical scale (and therefore very much looks like a {\em vacuum diagram} or a {\em vacuum bubble}), appears to be a quite natural tool to achieve, in practice, a separation between physical observables and vacuum fluctuations. 
 This idea is pictorially illustrated in  figure~\ref{fig:vacuum}, which represents a generic diagram contributing to a connected Green function.
When the loop (sub)-integration momenta become large, all momenta attached to it and all internal masses become negligible and the (sub)-diagram effectively behaves like a {\em vacuum bubble}, as shown in figures~\ref{fig:vacuum} (b) and (c), and decouples. Such large loop momentum states (which I label {\em vacuum configurations}), being universal, do not belong to the interaction any more and should be removed. 
In this paper, I adopt a very pragmatic approach and show that the integral over the terms which remain after subtracting these unphysical modes owns all good properties one wishes to perform practical calculations, namely it is four-dimensional, gauge invariant and independent on any cutoff, as will be explained in the next section. In this respect, one could just {\em define} differently the loop integrals, so that the mechanism nature uses to wipe the infinities out, either by reabsortion into the vacuum, as just shown, or via renormalization, becomes less relevant.
\section{The FDR integral}
\label{sec:fdr} 
The first problem is how to recognize, classify and subtract from any given diagram the unphysical large loop momentum configurations, and how to deal with them. Eqs.~\ref{eq:eqden1} and~\ref{eq:eqden2} serve as a guideline. First of all, a convenient parametrization is needed, in terms of an arbitrary scale, called $\mu$, compared to which the loop momenta are considered to be large. Thus, one expects the high frequencies to decouple in the limit $\mu \to 0$. Secondly, one requires independence on the UV cutoff, which means no left-over dependence on $\mu$ in physical quantities. With all of that in mind, I represent {\em vacuum configurations} as $\ell$-loop integrals which only depend on the unphysical scale $\mu$. 
A rank-r one-loop example is 
\begin{eqnarray}
\label{eq:ex1}
\mur^{-\epsilon} \int d^n q \frac{q_{\alpha_1} \cdots q_{\alpha_r}}{(q^2-\mu^2)^j}\,,
\end{eqnarray}
and a scalar two-loop case reads
\begin{eqnarray}
\label{eq:ex2}
\mur^{-2 \epsilon} \int d^n q_1 d^n q_2 \frac{1}{(q_1^2-\mu^2)^{j_1} (q_2^2-\mu^2)^{j_2} ({(q_1+q_2)}^2-\mu^2)^{j_3}}\,.
\end{eqnarray}
Note that, according to the values of $j$, $j_1$ $j_2$ and $j_3$, they may be or may be not UV divergent.
As a matter of definition, I call {\em vacuum integrands} (or {\em vacuum terms}) all integrands such as those appearing in eqs.~\ref{eq:ex1} and~\ref{eq:ex2} which only depend on the unphysical scale $\mu$ \footnote{Abusing a bit the language, I define finite, logarithmic and quadratic divergent vacuum integrands those generating finite, globally logarithmic and quadratic divergent integrals, etc.}.
Let now $I^{\rm DR}_{\ell-loop}$ a representative $\ell$-loop diagram contributing to a Green function computed in DR. Then
\begin{eqnarray}
I^{\rm DR}_{\ell-loop}= 
\mur^{-\ell \epsilon} \int \prod_{i=1}^{\ell} d^nq_{i} \,
J(\{q_{\alpha}, q^2 , \rlap/ q \})\,,
\end{eqnarray}
where $\{q_{\alpha}, q^2 , \rlap/ q \}$ symbolically denotes the set 
of loop-integration variables upon which the integrand $J$ depends, where I distinguish among tensor like structures, denoted by $q_{\alpha}$, self contracted loop-momenta $q^2$ and contractions with $\gamma$ matrices, $\rlap/ q$. 
Clearly, as done in eq.~\ref{eq:eqlim02} \footnote{I assume here no infrared or collinear divergences in
$I^{\rm DR}_{\ell-loop}$. Such cases are discussed in section~\ref{sec:icol}.}, 
\begin{eqnarray}
\label{eq:eq24}
I^{\rm DR}_{\ell-loop}= \lim_{\mu \to 0}\, 
\mur^{-\ell \epsilon} \int \prod_{i=1}^{\ell} d^nq_{i} \,
J(\{q_{\alpha}, q^2-\mu^2 , \rlap/ q - \mu \})\,,
\end{eqnarray}
where it is understood that the replacements in the integrand affect not only boson and fermion propagators, which get modified as follows \footnote{A small negative imaginary part of $\mu$ generates the correct $+ i 0$ propagator prescription as well.} 
\begin{eqnarray}
\label{eq:eq1}
\frac{1}{(q+p)^2-m^2} &\to& 
\frac{1}{(q+p)^2-m^2-\mu^2}\,,\nonumber \\
\frac{1}{\rlap/q +\rlap/p   -m} &\to& 
\frac{1}{\rlap/q +\rlap/p   -m-\mu} \,,
\end{eqnarray}
but also numerators, as required by gauge invariance.
In the following, $\mu$ will often appear in combination with the propagators $D_i$, $q^2$ and $\rlap/ q$, thus I use the same notation introduced in the previous section, namely
\begin{eqnarray}
\label{eq:eq1a}
\bar{D}_i     \equiv D_i-\mu^2\,,~~
\bar{q}^2    \equiv q^2-\mu^2\,,~~ 
\rlap/\bar{q} \equiv \rlap/q - \mu \,. 
\end{eqnarray}
Eq.~\ref{eq:eq24} is the place where the unphysical scale $\mu$ which separates the large-loop vacuum configurations is introduced. 
Suppose now to use eq.~\ref{eq:id} to split the integrand $J$ into a part containing all possible divergent vacuum integrands ($J_{V}$) and a piece integrable 
in four dimensions~\footnote{This splitting is similar to that one used in
\cite{Cherchiglia:2010yd}.}
\begin{eqnarray}
\label{eq:split}
  J(\{q_{\alpha}, \bar q^2 , \rlap/ \bar q \}) = 
J_V(\{q_{\alpha}, \bar q^2 , \rlap/ \bar q \})+  
J_F(\{q_{\alpha}, \bar q^2 , \rlap/ \bar q \})\,.
\end{eqnarray}
In order to keep gauge cancellations, it is important
to perform this separation globally on $\bar q^2$ and $\rlap/ \bar q$, meaning that their $\mu$ parts should not be treated differently from $q^2$ and $\rlap/ q$ \footnote{I call this a 
{\em global treatment} of $\bar q^2$ and $\rlap/ \bar q$.}. 
The FDR integral is then defined as
\begin{eqnarray}
\label{eq:intfdr}
I^{\rm FDR}_{\ell-loop} = \int \prod_{i=1}^{\ell} [d^4q_{i}] \,
J(\{q_{\alpha}, \bar q^2 , \rlap/ \bar q \})\,\equiv
\lim_{\mu \to 0}\, 
\left.\int \prod_{i=1}^{\ell} d^4q_{i} 
J_F(\{q_{\alpha}, \bar q^2 , \rlap/ \bar q \})\right|_{\mu = \mur}\,,
\end{eqnarray}
where the symbol ${\displaystyle \int [d^4q]}$ means: 
\begin{enumerate}
\item use eq.~\ref{eq:id} to move all divergences in vacuum integrands, treating
$\bar q^2$ and $\rlap/ \bar q$ globally;
\item drop all divergent vacuum terms from the integrand;
\item integrate over $d^4q$;
\item take the limit $\mu \to 0$, until a logarithmic dependence on $\mu$ 
is reached;
\item compute the result in $\mu = \mur$.
\end{enumerate}
The FDR integral is a physical quantity in which all high frequencies giving rise to unphysical vacuum configurations either do not contribute or are fully subtracted.
Consider, in fact, the connection of $I^{\rm FDR}_{\ell-loop}$ with the original integral of eq.~\ref{eq:eq24} 
\begin{eqnarray}
\label{eq:fdrconn}
I^{\rm FDR}_{\ell-loop} =  I^{\rm DR}_{\ell-loop} -
\lim_{\mu \to 0}\, 
\mur^{-\ell \epsilon} 
\left.\int \prod_{i=1}^{\ell} d^nq_{i} 
J_V(\{q_{\alpha}, \bar q^2 , \rlap/ \bar q \})\right|_{\mu = \mur}\,.
\end{eqnarray}
Since the only available scale is $\mu$, the contribution to $J_V$ of polynomially divergent vacuum integrands vanishes, when $\mu \to 0$. Thus, large loop
polynomially divergent vacuum configurations immediately decouple.
Logarithmically divergent integrands give instead a contribution of the form
\bqa
\label{eq:log}
K + \sum_{i=1}^{\ell} a_i \ln^i(\mu/\mur)\,,
\eqa
where $K$ contains infinities and constant terms.
Since $I^{\rm DR}_{\ell-loop}$ is independent on $\mu$, both sides of ~\ref{eq:fdrconn} have the same $\mu$ dependence given in eq.~\ref{eq:log}\,, before computing them at $\mu= \mur$. The powers of  $\ln(\mu)$, being generated by the low energy regime of the integration momenta, do not decouple and should be moved to the physical part $I^{\rm FDR}_{\ell-loop}$. Then one can take the formal limit $\mu \to 0$ in it \footnote{This is important because no dependence on the initial cutoff is allowed in the physical part.} by trading $\mu$ for $\mur$.
 Differently stated, the point $\mu=\mur$ \footnote{In which eq.~\ref{eq:log} reduces to $K$.} is such that also the logarithmic divergent vacuum bubbles completely decouple.
As for the high frequencies of finite vacuum bubbles, they naturally give 
vanishing contributions at large values of the loop momenta, by power counting. If one subtracts them, unphysical arbitrary powers of $1/\mu$ are generated in the physical part, that should be compensated by moving back to it analogous poles created in $J_V$.

Eq.~\ref{eq:intfdr} defines a multi-loop integral with all good properties one expects. It is finite in four dimensions, cut-off independent and invariant under any shift of the integration variables. The latter property follows from the fact that it can be also defined as the difference of two DR integrals, as in eq.~\ref{eq:fdrconn}.
It respects gauge invariance by construction, because of the shift invariance and of the {\em global treatment} of $\bar q^2$ and $\rlap/ \bar q$. This means that properties such as
\bqa
\label{eq:gi1}
\int \prod_{i=1}^{\ell} [d^4q_{i}] \,
J(\{q_{\alpha}, \bar q^2 , \rlap/ \bar q \}) &=&
 \int \prod_{i=1}^{\ell} [d^4q_{i}] \,
J(\{q_{\alpha}, \bar q^2 , \rlap/ \bar q \})\frac{\bar q_j^2}{\bar q_j^2-M^2} 
\nonumber \\
&-& \int \prod_{i=1}^{\ell} [d^4q_{i}] 
J(\{q_{\alpha}, \bar q^2 , \rlap/ \bar q \})\frac{M^2}{\bar q_j^2-M^2}
\,~~~
\forall q_j \in \{q_{\alpha}, \bar q^2 , \rlap/ \bar q \} \nonumber \\
\eqa
are guaranteed.
\section{One-loop examples}
\label{sec:1loop}
\subsection{Scalar integrals not depending on external momenta}
I start with a simple logarithmically divergent integral with no external scale
\begin{equation}
\label{eq:eq2}
 \int [d^4q] \frac{1}{\bar D^2}\,,
\end{equation}
with $\bar D$ given in eq.~\ref{eq:den}.
Eq.~\ref{eq:eqden2} with $p= 0$ can be used to extract 
the divergent vacuum integrands, which, from now on, I write inside square brackets
\begin{equation}
\label{eq:eq3}
\frac{1}{\bar{D}^2} = \left[\frac{1}{\bar{q}^4}\right]+M^2
\left(\frac{1}{{\bar D}^2 \bar{q}^2} + \frac{1}{{\bar D} \bar{q}^4} \right)\,.
\end{equation}
Then \footnote{See appendix~\ref{sec:appa}.}
\begin{equation}
\label{eq:eq4}
I_0^{\rm FDR} = \int [d^4q] \frac{1}{\bar{D}^2} 
\equiv 
\lim_{\mu \to 0 } M^2 \left. \int d^4q \left(\frac{1}{{\bar D}^2 \bar{q}^2} + \frac{1}{{\bar D} \bar{q}^4} \right)\right|_{\mu = \mur} 
= -i \pi^2\, \ln \frac{M^2}{\mur^2}\,.
\end{equation}
In an analogous way, eq.~\ref{eq:eqden1} gives \footnote{See again appendix~\ref{sec:appa}.}
\begin{equation}
\label{eq:eq7}
I_2^{\rm FDR} = \int [d^4q] \frac{1}{\bar D} \equiv
\lim_{\mu \to 0 } M^4 \left.\int d^4q \frac{1}{{\bar D} \bar{q}^4}\right|_{\mu=\mur}
=  -i \pi^2\,M^2\,\left(
\ln \frac{M^2}{\mur^2}-1
         \right)\,.
\end{equation}
Note also that, by definition of FDR
\begin{eqnarray}
\int[d^4q] \, (\bar{q}^2)^j = 0~~~\forall j \ge -2\,.
\end{eqnarray}
Eqs.~\ref{eq:eq4} and~\ref{eq:eq7} coincide with 
the corresponding dimensionally regulated expressions 
in the $\overline{\rm MS}$ scheme. 
This is because only one logarithmically divergent scalar integrand exists 
at one-loop, which can contribute to $J_V$ in eq.~\ref{eq:fdrconn}. Thus, a correspondence exists
\begin{eqnarray}
\label{eq:corr1l}
 \begin{tabular}{lcl}
$\displaystyle{\frac{1}{\epsilon}}$ + UC subtraction after 
integration & $\leftrightarrow$ &
$\displaystyle{\frac{1}{\bar{q}^4}}$ subtraction before integration.
\end{tabular} 
\end{eqnarray} 
Universal Constants (UC) also appear in the l.h.s. because the full 
integrand is subtracted in FDR.

Eq.~\ref{eq:corr1l} serves to show the formal equivalence, at one-loop, of DR
after renormalization and FDR. However, as already observed at the end of 
section~\ref{sec:sec1}, the important difference between the two approaches is that the former requires to renormalize away the divergences, while in the latter the subtraction is part of the definition of FDR integral, so that, by construction, no UV infinities appear from the very beginning. 
\subsection{Shifting the integration momentum}
Although shift invariance is guaranteed by construction, it is instructive to verify this property in simple cases.
Let me consider, for example, the shifted versions of $I_0^{\rm FDR}$ and $I_2^{\rm FDR}$
\begin{eqnarray} 
\label{eq:eq11}
I_{0p}^{\rm FDR} &=& \int [d^4q] \frac{1}{{\bar D}^2_p} \,,
 \nonumber \\
I_{2p}^{\rm FDR} &=& \int [d^4q] \frac{1}{{\bar D}_p}\,,
\end{eqnarray} 
where $\bar{D_p}= (q+p)^2-M^2-\mu^2$ and $p$ is an arbitrary 4-vector.
By iteratively using eq.~\ref{eq:id} one rewrites
\begin{eqnarray} 
\label{eq:eq12}
\frac{1}{{\bar D}^2_p} &=& \left[\frac{1}{{\bar q}^4}\right]
 +
 \frac{d(q)}{{\bar q}^4{\bar D}_p}
+\frac{d(q)}{{\bar q}^2{\bar D}^2_p} \,,
\nonumber \\
\frac{1}{{\bar D}_p} &=&
\left[
 \frac{1}{{\bar q}^2}
+\frac{d(q)}{{\bar q}^4}
+\frac{(d_1 \cdot q)^2}{{\bar q}^6}
\right]
 +
 \frac{d_0^2+2 d_0(d_1 \cdot q)}{{\bar q}^4{\bar D}_p}
+\frac{d(q)(d_1 \cdot q)^2}{{\bar q}^6{\bar D}_p} \,,
\end{eqnarray} 
with
\begin{eqnarray} 
\begin{tabular}{ll}
$d_0  =  M^2-p^2\,,$              & $d^{\mu}_1  =  -2 p^\mu\,,$ 
\end{tabular}
\end{eqnarray} 
which implies, by definition
\begin{eqnarray}
\label{eq:eq12a}
\int [d^4q] \frac{1}{{\bar D}^2_p} &\equiv&
\lim_{\mu \to 0 } \left.\int d^4q
 \left(
 \frac{d(q)}{{\bar q}^4{\bar D}_p}
+\frac{d(q)}{{\bar q}^2{\bar D}^2_p} 
\right)\right|_{\mu = \mur}\,, \nonumber \\
\int [d^4q] \frac{1}{{\bar D}_p} &\equiv& 
\lim_{\mu \to 0 } \left. \int d^4q
 \left(
 \frac{d_0^2+2 d_0(d_1 \cdot q)}{{\bar q}^4{\bar D}_p}
+\frac{d(q)(d_1 \cdot q)^2}{{\bar q}^6{\bar D}_p} 
\right)\right|_{\mu = \mur} \,.
\end{eqnarray}
Integrating the second of eqs.~\ref{eq:eq12a} is a bit involved, 
but there are not surprises:
\bqa
I_{0p}^{\rm FDR} &=& I_{0}^{\rm FDR}\,,
 \nonumber \\
I_{2p}^{\rm FDR} &=& I_{2}^{\rm FDR}\,.
\eqa
\subsection{Scalar integrals depending on external momenta}
Equipped with the previous results, it is straightforward to show that
\begin{eqnarray}  
\label{eq:eq13a}
\int [d^4q] \frac{1}{\bar{D}_0 \bar{D}_1} &\equiv&
\lim_{\mu \to 0}
\left. \int d^4q \left(
\frac{d(q)}{{\bar q}^4\bar{D}_1}
+m_0^2\frac{1}{{\bar q}^4\bar{D}_0}
+m_0^2\frac{d(q)}{{\bar q}^4\bar{D}_0\bar{D}_1}
\right)\right|_{\mu = \mur} \nonumber \\
 &=& -i \pi^2 \int_0^1 d \alpha \ln \frac{\chi(\alpha)}{\mur^2}\,,
\end{eqnarray}
where 
\begin{equation}
\bar{D}_0= q^2     -m_0^2-\mu^2\,,~~ 
\bar{D}_1= (q+p)^2 -m_1^2-\mu^2\,,~~
d(q)= m^2-p^2-2(p\cdot q)\,,
\end{equation}
and
\begin{equation}
\chi(\alpha)= m_0^2\alpha+m_1^2(1-\alpha)-p^2\alpha(1-\alpha)\,.
\end{equation}
Eq.~\ref{eq:eq13a} again coincides with the $\overline{\rm MS}$ result.
Notice also that, due to shift invariance,
\begin{equation}
\int [d^4q] \frac{1}{\bar{D}_0 \bar{D}_1} =
\int_0^1 d \alpha \int [d^4q] \frac{1}{[{\bar q}^2-\chi(\alpha)]^2}\,.
\end{equation}
\subsection{Tensor integrals}
From the identities 
\begin{eqnarray}
\label{eq:eq14a}
\frac{1}{{\bar D}^2}
 &=& \left[\frac{1}{{\bar q}^4}+2 \frac{M^2}{{\bar q}^6}\right]
          + M^4\left(\frac{2}{{\bar D}{\bar q}^6}
                    +\frac{1}{{\bar D}^2{\bar q}^4} 
               \right)\,,\nonumber \\
\frac{1}{{\bar D}^3}
 &=& \left[\frac{1}{{\bar q}^6} \right]
          + M^2\left(\frac{1}{{\bar D}^3{\bar q}^2}
                    +\frac{1}{{\bar D}^2{\bar q}^4} 
                    +\frac{1}{{\bar D}  {\bar q}^6} 
               \right)\,,\nonumber \\
\frac{1}{{\bar D}^4}
 &=& \left[\frac{1}{{\bar q}^8} \right]
          + M^2\left(\frac{1}{{\bar D}^4{\bar q}^2}
                    +\frac{1}{{\bar D}^3{\bar q}^4} 
                    +\frac{1}{{\bar D}^2{\bar q}^6} 
                    +\frac{1}{{\bar D}  {\bar q}^8} 
               \right)\,,\nonumber \\
\frac{1}{{\bar D}^3}
 &=& \left[\frac{1}{{\bar q}^6}+3 \frac{M^2}{{\bar q}^8} \right]
          + M^4\left(\frac{3}{{\bar D}  {\bar q}^8} 
                    +\frac{2}{{\bar D}^2{\bar q}^6} 
                    +\frac{1}{{\bar D}^3{\bar q}^4} 
               \right)\,,
\end{eqnarray}
with $\bar D$ defined in eq.~\ref{eq:den}, one obtains
\begin{eqnarray}
\label{eq:eq14}
\int [d^4q] 
\frac{q^\alpha q^\beta}{{\bar D}^2} 
 &=& M^4 \lim_{\mu \to 0}
 \left. \int d^4q\,q^\alpha q^\beta
\left(               \frac{2}{{\bar D}{\bar q}^6}
                    +\frac{1}{{\bar D}^2{\bar q}^4} 
               \right) \right|_{\mu=\mur}\,,
\nonumber \\
\int [d^4q] \frac{q^\alpha q^\beta}{{\bar D}^3} 
 &=& M^2 \lim_{\mu \to 0}
\left. \int d^4q \,q^\alpha q^\beta
\left(               \frac{1}{{\bar D}^3{\bar q}^2}
                    +\frac{1}{{\bar D}^2{\bar q}^4} 
                    +\frac{1}{{\bar D}  {\bar q}^6} 
               \right)\right|_{\mu=\mur} \,,
\nonumber \\
\int [d^4q] \frac{q^\alpha q^\beta q^\gamma q^\delta}{{\bar D}^4} 
 &=&  M^2 \lim_{\mu \to 0}
\left. \int d^4q\,q^\alpha q^\beta q^\gamma q^\delta
\left(               \frac{1}{{\bar D}^4{\bar q}^2}
                    +\frac{1}{{\bar D}^3{\bar q}^4} 
                    +\frac{1}{{\bar D}^2{\bar q}^6} 
                    +\frac{1}{{\bar D}  {\bar q}^8} 
               \right)\right|_{\mu=\mur}\,,
\nonumber \\
\int [d^4q] \frac{q^\alpha q^\beta q^\gamma q^\delta}{{\bar D}^3} 
 &=&  M^4 \lim_{\mu \to 0}
 \left. \int d^4q\,q^\alpha q^\beta q^\gamma q^\delta
\left(              \frac{3}{{\bar D}  {\bar q}^8} 
                    +\frac{2}{{\bar D}^2{\bar q}^6} 
                    +\frac{1}{{\bar D}^3{\bar q}^4} 
               \right)\right|_{\mu=\mur}\,.
\end{eqnarray}
Given the four-dimensional definition, one can replace 
\begin{eqnarray}
q^\alpha q^\beta               &\to& \frac{q^2}{4}  g^{\alpha \beta}\,,  \nonumber \\
q^\alpha q^\beta q^\gamma q^\delta &\to& \frac{q^4}{24} g^{\alpha \beta \gamma \delta}\,, 
\nonumber \\
g^{\alpha \beta \gamma \delta} &\equiv& (g^{\alpha \beta}  g^{\gamma \delta}
                                 +g^{\alpha \gamma} g^{\beta\delta}
                                 +g^{\alpha \delta}  g^{\beta\gamma}) \,, 
\end{eqnarray}
in eq.~\ref{eq:eq14} and compute
\begin{eqnarray}
\label{eq:eq15}
\int [d^4q] \frac{q^\alpha q^\beta}{{\bar D}^2} 
 &=& \frac{g^{\alpha \beta}}{2}  I_2^{\rm FDR}\,,
\nonumber \\
\int [d^4q] \frac{q^\alpha q^\beta}{{\bar D}^3} 
 &=&  \frac{g^{\alpha \beta}}{4}  I_0^{\rm FDR} \,,
\nonumber \\
\int [d^4q] \frac{q^\alpha q^\beta q^\gamma q^\delta}{{\bar D}^4} 
 &=&  \frac{g^{\alpha \beta \gamma \delta}}{24}  I_0^{\rm FDR} \,,
\nonumber \\
\int [d^4q] \frac{q^\alpha q^\beta q^\gamma q^\delta}{{\bar D}^3} 
 &=& \frac{g^{\alpha \beta \gamma \delta}}{8}  I_2^{\rm FDR}\,,
\end{eqnarray}
with $I_0^{\rm FDR}$ and $I_2^{\rm FDR}$ given in  eqs.~\ref{eq:eq4} and~\ref{eq:eq7}.
Finally, odd-rank vacuum tensor integrals, such as
\begin{eqnarray}
\int [d^4q] \frac{q^{\alpha_1} q^{\alpha_2} \cdots q^{\alpha_{(2n+1)}}}{{\bar D}^m}\,,
\end{eqnarray}
vanish due to Lorentz invariance.
Eqs.~\ref{eq:eq15} are the gauge preserving conditions for one-loop tensor integrals~\cite{Wu:2003dd} which corroborate the consistency of the FDR approach.

Depending on the way a calculation is performed,
additional FDR integrals with powers of $\mu$ in the numerator may arise,
which should be treated accordingly to the {\em global treatment} prescription. 
For example
\begin{eqnarray}
\int [d^4q] \frac{\mu^2}{{\bar D}^3}\,.
\end{eqnarray}
should be considered as a logarithmically divergent one. 
FDR then requires an expansion of  $\frac{1}{{\bar D}^3}$ as 
in eq.~\ref{eq:eq14a}, giving rise to
\begin{eqnarray}
\label{eq:eq16}
\int [d^4q] \frac{\mu^2}{{\bar D}^3}
 =  M^2  \lim_{\mu \to 0} \mu^2
     \left. \int d^4q \left(\frac{1}{{\bar D}^3{\bar q}^2}
                    +\frac{1}{{\bar D}^2{\bar q}^4} 
                    +\frac{1}{{\bar D}  {\bar q}^6} 
               \right) \right|_{\mu = \mur} = \frac{i \pi^2}{2}\,.
\end{eqnarray}
Analogously
\begin{eqnarray}
\label{eq:eq17}
\int [d^4q] \frac{\mu^2}{{\bar D}^2} 
 &=& M^4 \lim_{\mu \to 0} \mu^2
\left. \int d^4q\,
\left(               \frac{2}{{\bar D}{\bar q}^6}
                    +\frac{1}{{\bar D}^2{\bar q}^4} 
               \right)  \right|_{\mu = \mur}= i \pi^2 M^2\,,
\nonumber \\
\int [d^4q] \frac{q^2\mu^2}{{\bar D}^4} 
 &=&  M^2 \lim_{\mu \to 0} \mu^2
\left. \int d^4q\,q^2
\left(               \frac{1}{{\bar D}^4{\bar q}^2}
                    +\frac{1}{{\bar D}^3{\bar q}^4} 
                    +\frac{1}{{\bar D}^2{\bar q}^6} 
                    +\frac{1}{{\bar D}  {\bar q}^8} 
               \right)  \right|_{\mu = \mur}= \frac{i \pi^2}{3}\,,
\nonumber \\
\int [d^4q] \frac{\mu^4}{{\bar D}^4} 
 &=&  M^2 \lim_{\mu \to 0} \mu^4
\left. \int d^4q\,
\left(               \frac{1}{{\bar D}^4{\bar q}^2}
                    +\frac{1}{{\bar D}^3{\bar q}^4} 
                    +\frac{1}{{\bar D}^2{\bar q}^6} 
                    +\frac{1}{{\bar D}  {\bar q}^8} 
               \right) \right|_{\mu = \mur}= -\frac{i \pi^2}{6}\,.
\nonumber \\
\end{eqnarray}
Eqs.~\ref{eq:eq16} and~\ref{eq:eq17} can be proved by direct 
integration. However, it is more elegant to observe that finite contributions 
may arise only when ${\mu^2}$ and ${\mu^4}$ hit $1/\mu^2$ and $1/\mu^4$ poles, respectively, which are more easily extracted by reinserting back
eq.~\ref{eq:eq14a}. Therefore
\begin{eqnarray}
\int [d^4q] \frac{\mu^2}{{\bar D}^3} &=& 
-\mu^2\int d^4q \frac{1}{{\bar q}^6}\,,\nonumber\\
\int [d^4q] \frac{\mu^2}{{\bar D}^2}   &=&
-2 M^2 \int d^4q \frac{1}{{\bar q}^6}\,,
\nonumber\\
\int [d^4q] \frac{q^2\mu^2}{{\bar D}^4} &=&
-\mu^2 \int d^4q \frac{q^2}{{\bar q}^8}\,,
\nonumber\\
\int [d^4q] \frac{\mu^4}{{\bar D}^4}    &=& 
-\mu^4 \int d^4q \frac{1}{{\bar q}^8}\,,
\end{eqnarray}
which reproduce the expected result.
Finally, by power counting
\begin{eqnarray}
\label{eq:eq17a}
\int [d^4q] \frac{\mu^{2j}}{\bar D^k} &=& 0\,~~{\rm when}~~k > 2+j\,, \nonumber\\\int [d^4q] \frac{\mu^{2j+1}}{\bar D^k} &=& 0\,.
\end{eqnarray}
It is quite remarkable the perfect parallelism between 
eqs.~\ref{eq:eq16},~\ref{eq:eq17} and~\ref{eq:eq17a} 
and their counterparts in DR
(see~\cite{Pittau:1996ez,Ossola:2007bb})
\begin{eqnarray}
\label{eq:eq18}
 \int [d^4q] \frac{\mu^2}{{\bar D}^3}   &=& 
-\int  d^nq  \frac{\tilde{q}^2}{{D}^3} \,,
\nonumber\\
 \int [d^4q] \frac{\mu^2}{{\bar D}^2}   &=&
-\int  d^nq  \frac{\tilde{q}^2}{{D}^2} \,,
\nonumber\\
\int [d^4q] \frac{q^2\mu^2}{{\bar D}^4} &=&
-\int  d^nq \frac{q^2\tilde{q}^2}{{D}^4} \,,
\nonumber\\
\int [d^4q] \frac{\mu^4}{{\bar D}^4}    &=& 
\int  d^nq \frac{\tilde{q}^4}{{D}^4}   \,,
\nonumber\\
\int [d^4q] \frac{\mu^{2j}}{\bar D^k}      &=&
\int  d^nq  \frac{{(\tilde{q}^2)}^{j}}{{D}^k} = 0\,~~{\rm when}~~k > 2+j\,,
\end{eqnarray}
where $\tilde{q}^2$ is the $\epsilon$-dimensional part of $q^2$.

With the help of these results, properties such
as \footnote{They can also be directly proved from the identities
in eq.~\ref{eq:eq14a}.}
\begin{eqnarray}
\label{eq:gi2}
\int [d^4q] \frac{\bar{q}^2-M^2}{\bar{D}^3} &=&  
\int [d^4q] \frac{1}{\bar{D}^2}\,,\nonumber \\
\int [d^4q] \frac{(\bar{q}^2-M^2)^2}{\bar{D}^4}  &=&
\int [d^4q] \frac{1}{\bar{D}^2}\,,\nonumber \\
\int [d^4q] \frac{\bar{q}^2-M^2}{\bar{D}^2} &=&  
\int [d^4q] \frac{1}{\bar{D}}\,, \nonumber \\
\int [d^4q] \frac{\bar{q}^2-M^2}{\bar{D}} &=&  
\int [d^4q] = 0\,,
\end{eqnarray}
follow, which guarantee that all usual manipulations 
are allowed in the FDR integrands.  
As a by-product, the Passarino-Veltman~\cite{Passarino:1978jh}, 
the OPP reduction approaches~\cite{Ossola:2006us,Ossola:2007ax},
together with all available literature to compute the contributions generated by the appearance of $\mu$ in the numerator
~\cite{Pittau:2011qp,Ossola:2008xq,Garzelli:2009is,Draggiotis:2009yb,Garzelli:2010qm,Shao:2011tg,Shao:2012ja}, can also be used in FDR.
\section{Two loops and beyond}
\label{sec:2loop}
As a two-loop example, consider the integral
\begin{equation}
I^{\rm FDR} = \int [d^4q_1][d^4q_2] \frac{1}{\bar D_1\bar D_2\bar D_{12}}\,,
\end{equation}
where
\begin{eqnarray}
\label{eq:2loop}
\bar D_1   &=& \bar{q}_1^2-m_1^2\,,\nonumber \\ 
\bar D_2   &=& \bar{q}_2^2-m_2^2\,,\nonumber \\
\bar D_{12} &=& \bar{q}_{12}^2-m_{12}^2\,,
\end{eqnarray}
and $q_{12} \equiv q_1 + q_2$.
Identity~\ref{eq:id} can be used to rewrite 
\begin{eqnarray}
\label{eq:eq21}
\frac{1}{\bar D_1\bar D_2\bar D_{12}} &=&
\left[\frac{1}{\bar{q}_1^2\bar{q}_2^2 \bar{q}_{12}^2}\right] \nonumber \\
&+&\frac{m_1^2}{(\bar D_1 \bar{q}_1^2)\bar{q}_2^2 \bar{q}_{12}^2}
+\frac{m_2^2}{\bar{q}_1^2 (\bar D_2 \bar{q}_2^2) \bar{q}_{12}^2}
+\frac{m_{12}^2}{\bar{q}_1^2 \bar{q}_{2}^2 (\bar D_{12} \bar{q}_{12}^2)}
\nonumber \\
&+& 
 \frac{m_1^2 m_2^2}{(\bar D_1 \bar{q}_1^2)(\bar D_2 \bar{q}_2^2)\bar{q}_{12}^2} 
+\frac{m_1^2 m_{12}^2}{(\bar D_1 \bar{q}_1^2)\bar{q}_{2}^2(\bar D_{12} \bar{q}_{12}^2)} 
+\frac{m_2^2 m_{12}^2}{\bar{q}_{1}^2(\bar D_2 \bar{q}_2^2)(\bar D_{12} \bar{q}_{12}^2)} 
\nonumber \\
&+& 
\frac{m_1^2 m_2^2 m_{12}^2}{(\bar D_1 \bar{q}_1^2)(\bar D_2 \bar{q}_2^2)(\bar D_{12} \bar{q}_{12}^2)}\,. 
\end{eqnarray}
The term between square brackets is, as usual, a vacuum integrand, 
which extracts the overall quadratic UV divergence of $I^{\rm FDR}$.
The following three produce logarithmically divergent (sub)-integrals and the last four can be integrated in four dimensions.
The next step is singling out the remaining divergences.
By rewriting
\begin{eqnarray}
\frac{1}{\bar{q}_{12}^2} &=& \frac{1}{\bar{q}_2^2}
             -\frac{q_1^2+2(q_1 \cdot q_2)} {\bar{q}_2^2\bar{q}_{12}^2}\,,
\end{eqnarray}
one obtains
\begin{eqnarray}
\label{eq:eq21a}
\frac{m_1^2}{(\bar D_1 \bar{q}_1^2)\bar{q}_2^2 \bar{q}_{12}^2} =
m_1^2
\left[
\frac{1}{\bar{q}_1^4\bar{q}_2^2 \bar{q}_{12}^2}
 \right]
+
 \frac{m_1^4}{(\bar D_1\bar{q}_1^4)}
\left[\frac{1}{\bar{q}_2^4} \right]
- m_1^4 \frac{q_1^2+2(q_1 \cdot q_2)}{(\bar D_1 \bar{q}_1^4)\bar{q}_2^4 \bar{q}_{12}^2}\,,
\end{eqnarray}
where the two vacuum integrands extract overall logarithmic
and overlapping logarithmic sub-divergences, respectively
\footnote{The structure of one-loop counterterm naturally appears for the latter.}, and the last term is integrable. The remaining two integrands in the second line of eq.~\ref{eq:eq21} can be treated analogously.
$I^{\rm FDR}$ then  reads
\begin{eqnarray}
\label{eq:eq22}
I^{\rm FDR} &\equiv& \lim_{\mu \to 0}
\int d^4q_1 \int d^4q_2
\left(
 \frac{m_1^2 m_2^2}{(\bar D_1 \bar{q}_1^2)(\bar D_2 \bar{q}_2^2)\bar{q}_{12}^2} 
+\frac{m_1^2 m_{12}^2}{(\bar D_1 \bar{q}_1^2)\bar{q}_{2}^2(\bar D_{12} \bar{q}_{12}^2)} 
+\frac{m_2^2 m_{12}^2}{\bar{q}_{1}^2(\bar D_2 \bar{q}_2^2)(\bar D_{12} \bar{q}_{12}^2)} 
\right.
\nonumber \\
&-& 
  m_1^4 \frac{q_1^2+2(q_1 \cdot q_2)}{(\bar D_1 \bar{q}_1^4)\bar{q}_2^4 \bar{q}_{12}^2}
- m_2^4 \frac{q_2^2+2(q_1 \cdot q_2)}{\bar{q}_1^4(\bar D_2 \bar{q}_2^4) \bar{q}_{12}^2} 
- m_{12}^4 \frac{q_{12}^2-2(q_1 \cdot q_{12})}{\bar{q}_1^4 \bar{q}_{2}^2
(\bar D_{12} \bar{q}_{12}^4)}
\nonumber \\
&+&
\left. \left.
\frac{m_1^2 m_2^2 m_{12}^2}{(\bar D_1 \bar{q}_1^2)(\bar D_2 \bar{q}_2^2)(\bar D_{12} \bar{q}_{12}^2)}
\right)\right|_{\mu = \mur}\,.
\end{eqnarray}
Other divergent two-loop integrals, such as
\begin{equation}
\int [d^4q_1][d^4q_2] \frac{1}{\bar D_1^2\bar D_2\bar D_{12}}\,,
\end{equation}
can be obtained by derivation with respect to masses.
 
\begin{figure}
\begin{center}
\begin{picture}(300,60)(0,0)

\SetOffset(230,15)
\BCirc(-30,12){20}
\CCirc(-50,12){2}{}{}
\Text(-20.5,12)[l]{$\mu$}

\SetOffset(130,15)
\BCirc(-30,12){20}
\CCirc(-50,12){2}{}{}
\Line(-30,32)(-30,-8)
\Text(-55,27)[l]{$\mu$}
\Text(-20.5,12)[l]{$\mu$}
\Text(-40,12)[l]{$\mu$}
\end{picture}
\caption{\label{fig:fig3} Two-loop (left) and one-loop (right) 
logarithmically divergent subtraction vacuum scalar integrands.
Dots denote propagator squared and $\mu$ is the unphysical 
mass running in each line.}

\end{center}
\end{figure}
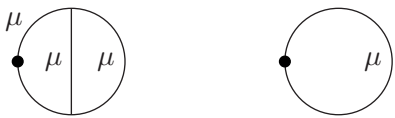

As already observed, quadratically divergent vacuum integrands, such as the first term in eq.~\ref{eq:eq21}, do not contribute when $\mu \to 0$, and one is left with the only two possible logarithmically
divergent subtraction scalar (sub)-diagrams shown 
in figure~\ref{fig:fig3}, which
are of the the type of those appearing in eq.~\ref{eq:eq21a}.
Therefore, besides eq.~\ref{eq:corr1l}, a two-loop correspondence holds
 \begin{eqnarray} 
\label{eq:corr2l}
\begin{tabular}{lcl}
$\displaystyle{\frac{1}{\epsilon^2}}$ + UC subtraction after 
integration & $\leftrightarrow$ &
$\displaystyle{\frac{1}{\bar{q}_1^4\bar{q}_2^2 \bar{q}_{12}^2}}$ subtraction before integration.
\end{tabular}  \nonumber \\
\end{eqnarray} 
Only Universal Constants appear in the left part of eq.~\ref{eq:corr2l} because of the existence of only one possible subtraction term.
As in the one-loop case, eq.~\ref{eq:corr2l} proves the equivalence, at two-loop, of DR after renormalization and FDR. The conceptual difference is again that, in FDR, there is no need to prove, for example, that overlapping divergences cancel, because all kinds of UV infinities are directly eliminated in the definition of FDR integral.

\begin{figure}
\begin{center}
\begin{picture}(300,128)(0,0)

\SetOffset(100,83)
\BCirc(-30,12){20}
\Line(-30,32)(-30,12)
\Line(-30,12)(-15,-1)
\Line(-30,12)(-45,-1)

\SetOffset(180,83)
\BCirc(-30,12){20}
\Line(-40,30)(-40,-6)
\Line(-20,30)(-20,-6)

\SetOffset(260,83)
\BCirc(-30,12){20}
\CCirc(-50,12){2}{}{}
\Line(-30,18)(-30,-8)
\BCirc(-30,25){7}


\SetOffset(140,18)
\BCirc(-30,12){20}
\CCirc(-50,12){2}{}{}
\GOval(-30,12)(10,19)(90){1}
\CCirc(-40,12){2}{}{}

\SetOffset(220,18)
\BCirc(-30,12){20}
\GOval(-30,12)(10,19)(90){1}
\CCirc(-49,18){2}{}{}
\CCirc(-49,6){2}{}{}

\end{picture} 
\caption{\label{fig:fig4} Irreducible three-loop topologies giving rise to logarithmically divergent subtraction vacuum diagrams. One dot denotes propagator squared; two dots mean propagator to the third power.}
\end{center}
\end{figure}
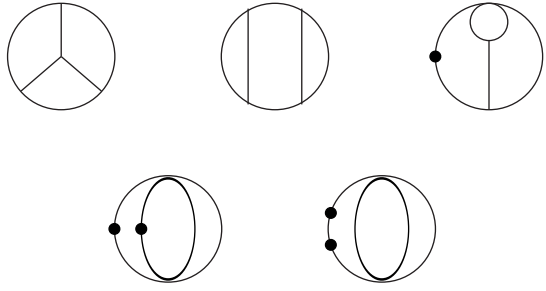

At three loop the situation is more involved because five irreducible topologies (see figure~\ref{fig:fig4}) may generate logarithmically divergent 
vacuum diagrams. In this case, FDR and DR might start differing 
diagram by diagram, although the gauge invariance properties of the FDR integral guarantee the equivalence of the two approaches \footnote{See section~\ref{sec:ren} for a more detailed discussion on this point.}.

\section{The ABJ anomaly and $\gamma_5$}
\label{sec:abj}
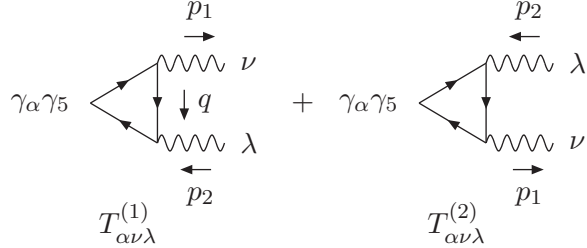
\begin{figure}
\begin{center}
\begin{picture}(300,100)(0,0)

\SetOffset(100,55)

\Text(-70,0)[l]{$ {\gamma_{\alpha}\gamma_5} $}
\ArrowLine(-40,0)(-15,15)
\ArrowLine(-15,-15)(-40,0)
\ArrowLine(-15,15)(-15,-15)

\Photon(-15,15)(10,15){3}{4}
\LongArrow(-5,25)(5,25)
\Text(-4,35)[l]{$ p_1 $}
\Text(16,15)[l]{$ {\nu} $}

\LongArrow(-5,5)(-5,-5)
\Text(0,0)[l]{$ q $}

\Photon(-15,-15)(10,-15){3}{4}
\LongArrow(5,-25)(-5,-25)
\Text(-4,-35)[l]{$ p_2 $}
\Text(16,-15)[l]{$ {\lambda} $}
\Text(-15,-45)[r]{$ {T^{(1)}_{\alpha\nu\lambda}}$}

\Text(35,0)[l]{$ + $}

\SetOffset(223,55)

\Text(-70,0)[l]{$ {\gamma_{\alpha}\gamma_5} $}
\ArrowLine(-40,0)(-15,15)
\ArrowLine(-15,-15)(-40,0)
\ArrowLine(-15,15)(-15,-15)

\Photon(-15,15)(10,15){3}{4}
\LongArrow(5,25)(-5,25)
\Text(-4,35)[l]{$ p_2 $}
\Text(16,15)[l]{$ {\lambda} $}

\Photon(-15,-15)(10,-15){3}{4}
\LongArrow(-5,-25)(5,-25)
\Text(-4,-35)[l]{$ p_1 $}
\Text(16,-15)[l]{$ {\nu} $}
\Text(-15,-45)[r]{$ {T^{(2)}_{\alpha\nu\lambda}}$}

\end{picture}
\caption{\label{fig:fig1} The two diagrams generating the ABJ anomaly.}
\end{center}
\end{figure}

In this section, I reproduce the known ABJ 
anomaly~\cite{Adler:1969gk,Bell:1969ts} with the FDR approach. 
Two massless fermion loop diagrams contribute, as shown 
in figure~\ref{fig:fig1}.  
When contracted with $p= p_1-p_2$, the first term gives
\begin{equation}
p^\alpha T^{(1)}_{\alpha \nu \lambda}=-i \frac{e^2}{(2 \pi)^4}
\int [d^4q] \,
 {\rm Tr}\left[\rlap/p \gamma_5 
\frac{1}{\rlap/Q_2-\mu} \gamma_\lambda
\frac{1}{\rlap/Q_0-\mu} \gamma_\nu
\frac{1}{\rlap/Q_1-\mu}
 \right]\,,
\end{equation}
where $Q_i= q+p_i$ ($p_0 = 0$) and all massless propagators 
are shifted by $\mu$, as required by FDR.
Rewriting
\begin{equation}
\label{eq:eq19}
\rlap/p = (\rlap/Q_1-\mu) - (\rlap/Q_2-\mu)
\end{equation}
produces
\begin{eqnarray}
p^\alpha T^{(1)}_{\alpha \nu \lambda}&=&-i \frac{e^2}{(2 \pi)^4}
\int [d^4q] 
\left(
 {\rm Tr}\left[\gamma_5 
\frac{1}{\rlap/Q_2-\mu} \gamma_\lambda
\frac{1}{\rlap/Q_0-\mu} \gamma_\nu
 \right]
-{\rm Tr}\left[\gamma_5 
\frac{1}{\rlap/Q_{-1}-\mu} \gamma_\nu
\frac{1}{\rlap/Q_0-\mu} \gamma_\lambda
 \right]
\right.
 \nonumber \\
&+&2 \mu
\left. 
 {\rm Tr}\left[\gamma_5 
\frac{1}{\rlap/Q_2-\mu} \gamma_\lambda
\frac{1}{\rlap/Q_0-\mu} \gamma_\nu
\frac{1}{\rlap/Q_1-\mu}
 \right]
\right)\,,
\end{eqnarray}
where a shift $q \to q-p_1$ \footnote{Legal in FDR.} has been performed in the second trace, and $Q_{-1} \equiv q -p_1$. The contribution of the second diagram
is obtained by replacing $p_1 \leftrightarrow -p_2$ and $\lambda \leftrightarrow \nu$, thus all terms not proportional to $\mu$ drop in the sum
$T= T^{(1)}+T^{(2)}$
\begin{eqnarray}
p^\alpha T_{\alpha \nu \lambda}
= -i \frac{e^2}{4\pi^4} 
{\rm Tr}[\gamma_5 \rlap/p_2 \gamma_\lambda \gamma_\nu \rlap/p_1]
\int [d^4q] \,
\mu^2
 \frac{1}{\bar D_0 \bar D_1 \bar D_2}\,,
\end{eqnarray}
where ${\bar D_i} = (q+p_i)^2-\mu^2$.
The FDR integral is given in eq.~\ref{eq:eq16}, leading to
the correct answer
\begin{eqnarray}
\label{eq:eq20}
p^\alpha T_{\alpha \nu \lambda}
= \frac{e^2}{8\pi^2} 
{\rm Tr}[\gamma_5 \rlap/p_2 \gamma_\lambda \gamma_\nu \rlap/p_1]\,.
\end{eqnarray}

\begin{figure}
\begin{center}
\begin{picture}(300,80)(0,0)
\SetOffset(180,35)

\Photon(-80,0)(-40,0){3}{5}
\Photon(-40,0)(-15,15){3}{5}
\Photon(-15,-15)(-40,0){3}{5}
\ArrowLine(-15,-15)(-15,15)
\ArrowLine(-15,15)(10,27)
\ArrowLine(10,-27)(-15,-15)
\Text(-25,24)[l]{$ {\gamma_5} $}

\end{picture}
\caption{\label{fig:fig2} Example of diagram in which the 
$\mu$ dependence cannot be fully reabsorbed into the fermion masses.}

\end{center}
\end{figure}
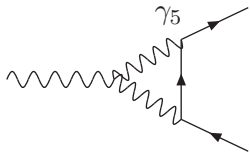

Finally, I comment on the role of $\gamma_5$.
In closed fermion loops, the FDR interpretation of $\mu$ as a propagator shift allows one to fully reabsorbe it into the fermion masses, providing unambiguous results also in the presence of $\gamma_5$. This was actually done in the previous calculation, where the result in eq.~\ref{eq:eq20} could also be obtained by starting with a physical fermion mass $m_f$ and replacing $m_f \to m_f + \mu$ in
the {\em finite} contribution proportional to $m_f^2$.   
However, in open fermion chains where not all $\rlap/q$ are linked to fermion masses, the ambiguity of {\em when} shifting $\rlap/q \to \rlap/q - \mu$ is potentially present. For example, in the combination
$$
( \cdots \rlap/ q \gamma_5 \frac{1}{\rlap/q} \cdots )\,
$$
occurring in the diagram of figure~\ref{fig:fig2}, the operations 
of anticommuting $\gamma_5$ and shifting $\rlap/q$ do not commute.
This is solved by considering the chiral theory as a good, gauge invariant starting point~\cite{Jegerlehner:2000dz}, which means, in practice, that $\gamma_5$ must be anticommuted towards the external spinors {\em before} shifting $\rlap/q$.
\section{Renormalization}
\label{sec:ren}
Since the UV infinities are subtracted right from the beginning, there is no need, in FDR, to add counterterms to the Lagrangian. One can indeed prove that
\begin{eqnarray}
\label{eq:eq25}
G^{\rm FDR}_{\ell-loop}(\mur) = G^{\rm DR}_{\ell-loop}(\mur)\,,   
\end{eqnarray}
where $G^{\rm FDR}$ is a generic $\ell$-loop Green function computed in FDR, 
$G^{\rm DR}$ the same Green function calculated in DR, but after renormalization \footnote{I assume the same renormalized parameters in $G^{\rm DR}$ and $G^{\rm FDR}$.}, and $\mur$ is the renormalization scale.
The generic form of $G^{\rm FDR}$ is
\begin{equation}
\label{eq:eq23}
G^{\rm FDR}_{\ell-loop}(\mur) = \sum_{i=0}^{\ell} a^{\rm DR}_i \log^i(\mur)
+ {\rm R}^{\rm DR}(\{p,M\}) + {\rm R_0} \,, 
\end{equation}
where ${\rm R}^{\rm DR}(\{p,M\})$ is a term depending on the kinematical variables of the process and ${\rm R_0}$ an a-dimensional constant.
By construction, the coefficients $a^{\rm DR}_i$ and 
${\rm R}^{\rm DR}(\{p,M\})$ are the same one would compute in DR. In fact, the logarithmic dependence on $\mur$ is fully included in the definition of FDR integral, and any kinematical information is only contained in the finite
part $J_F$ of eq.~\ref{eq:split}, which is common to both DR and FDR.
The only possible difference is ${\rm R}_0$, because FDR requires to entirely drop the constant parts of the logarithmically divergent integrals, while DR only
infinities and Universal Constants.
However, one proves, by contradiction, that also ${\rm R_0}$ is common in 
the two schemes.
If FDR misses some part of ${\rm R_0}$, it could be fixed back by enforcing the Ward-Slavnov-Taylor identities of the theory, which requires a computation in FDR in order to be able to adjust, by hand, all terms which violate them. But since FDR respects, by construction, all cancellations needed to prove
gauge invariance (such as eqs.~\ref{eq:gi1} and \ref{eq:gi2}) no violation would be found, therefore eq.~\ref{eq:eq25} is proven.

If $G^{\rm FDR}_{\ell-loop}$ belongs to a renormalizable theory, expressing its free parameters in terms of observables trades the unphysical scale $\mur$ for a physical one, ensuring the equivalence of FDR with any other consistent renormalization scheme. Particularly interesting is the case of non-renormalizable field theories, such as quantum gravity. Measuring the parameters of the Lagrangian does not guarantee any longer the disappearance of $\mur$. However, $G^{\rm FDR}_{\ell-loop}$ is finite in four dimensions and eq.~\ref{eq:eq23} still holds, so that one additional measurement can be used, in principle, to fix $\mur$ order by order, making the theory predictive
\footnote{This is allowed because the limit $\mu \to 0$ is fully taken in 
$G^{\rm FDR}_{\ell-loop}$, and, as a result, the original $\mu$ is replaced by $\mur$, which does not need to be small.}. A possible interpretation is that the original theory, possibly due to non perturbative effects, or to a very complicate structure of its vacuum, could be not complete enough, or not to allow, a full determination of its UV counterpart, as happens in renormalizable theories. If it is the case, eq.~\ref{eq:eq23} could still provide a parametrization, in terms of $\mur$, of unknown unphysical phenomena which do not decouple. 
 Measuring $\mur$, definitively integrates out all unphysical degrees of freedom, leaving the observable spectrum free of UV effects.
Whether this is a valid way out, is debatable. The important points to keep in mind are that
\begin{itemize} 
\item the Lagrangian is left untouched;
\item gauge invariance is not broken; 
\item four-dimensionality is kept;
\item one additional measurement is enough to fix $\mur$ at any perturbative order.
\end{itemize}
  
\section{Infrared and collinear divergences}
\label{sec:icol}
In massless theories, the FDR insertion of $\mu$ in the propagators (eq.~\ref{eq:eq1}) naturally regulates infrared and collinear divergences occurring in virtual loop integrals. This suggests the possibility to use FDR to regulate them also in the real emission.  In this section, I compute, as a one-loop 
example, the ${\cal O}(\alpha)$ QED corrections to the decay rate 
$\Gamma(Z \to f \bar f)$ with massless fermions, which provides an explicit example of FDR calculation in which all three types of divergences (UV, infrared and collinear) are simultaneously present. 

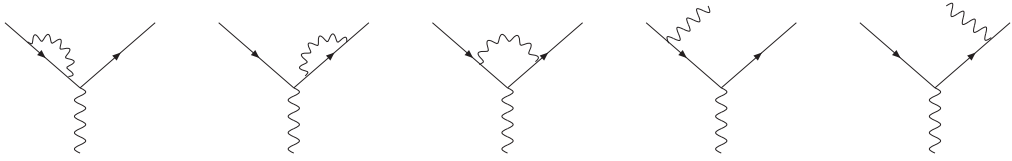
\begin{figure}
\begin{center}
\begin{picture}(450,90)(0,0)

\SetScale{0.7}
\SetOffset(50,47)
\ArrowLine(-30,35)(10,0)
\PhotonArc(-10,12)(15,-23,122){2}{6}
\ArrowLine(10,0)(50,35)
\Photon(10,0)(10,-35){3}{4}

\SetOffset(130,47)
\ArrowLine(-30,35)(10,0)
\ArrowLine(10,0)(50,35)
\PhotonArc(30,12)(15,57,202){2}{6}
\Photon(10,0)(10,-35){3}{4}

\SetOffset(210,47)
\ArrowLine(-30,35)(10,0)
\ArrowLine(10,0)(50,35)
\Photon(10,0)(10,-35){3}{4}
\PhotonArc(10,12)(15,8,178){2}{6}

\SetScale{0.7}
\SetOffset(290,47)
\ArrowLine(-30,35)(10,0)
\Photon(-18,24)(3,44){3}{4}
\ArrowLine(10,0)(50,35)
\Photon(10,0)(10,-35){3}{4}

\SetOffset(370,47)
\ArrowLine(-30,35)(10,0)
\ArrowLine(10,0)(50,35)
\Photon(40,27)(16,44){3}{4}
\Photon(10,0)(10,-35){3}{4}

\end{picture}
\caption{\label{fig:fig5} Virtual and real diagrams contributing 
to $Z \to f \bar f$.}
\end{center}
\end{figure}

By considering both virtual and real contributions as particular cuts of a 
two-loop FDR integral, in which the replacements
\bqa
\rlap/{q_j} &\to& \rlap/{q_j} -\mu \nonumber \\
q^2_j      &\to& q^2_j -\mu^2~~~~({\rm j= 1,2,12~as~in~eq.~\ref{eq:2loop}}) 
\eqa
are consistently performed both in numerators and denominators, causality requires 
\bqa
\frac{1}{q_j^2-\mu^2} \to \delta(q_j^2-\mu^2)\theta(q_j(0))
\eqa
for the cut lines.
One then expects the $\mu$ dependence to cancel when adding all contributing diagrams of figure~\ref{fig:fig5}, where $\mu$ is present, in the intermediate stages of the calculation, both in matrix elements and phase space integrals.

The virtual part can be computed very much as in the DR case, but the tensor reduction should be performed in four dimensions and the integrals interpreted in FDR. The needed loop functions are
\bqa
\label{eq:virtint}
B(s) &=&  \left. \int [d^4q] \frac{1}{(q^2-\mu^2)((q+p)^2-\mu^2)}\right|_{p^2= s} \,, \nonumber \\
B_0 &=&  \left. \int [d^4q] \frac{1}{(q^2-\mu^2)(q^2+2(q \cdot p))}\right|_{p^2= \mu^2} \,, \nonumber \\
B_1 &=&  \left.\frac{1}{p^2}\int [d^4q] \frac{(q \cdot p)}{(q^2-\mu^2)(q^2+2(q \cdot p))}\right|_{p^2= \mu^2} \,,  \nonumber \\
B^\prime_0 &=& \left. \frac{d}{d p^2}B_0 \right|_{p^2= \mu^2} \,,  \nonumber \\
B^\prime_1 &=& \left. \frac{d}{d p^2}B_1 \right|_{p^2= \mu^2} \,,  \nonumber \\
C(s)        &=& \left. \int [d^4q] 
\frac{1}{(q^2-\mu^2)(q^2+2(q \cdot p_1))(q^2-2(q \cdot p_2))}
\right|_{p_1^2= p_2^2 =\mu^2;(p_1+p_2)^2= s} \,, \nonumber \\
C_{\rm R}  &=& \int [d^4q]  
\frac{\mu^2}{(q^2-\mu^2)(q^2+2(q \cdot p_1))(q^2-2(q \cdot p_2))}\,, 
\eqa
and are listed in appendix~\ref{sec:appb}.  
The total virtual contribution reads
\begin{equation}
\label{eq:virtres}
\Gamma_{\rm V}(Z \to f \bar f)= \Gamma_0(Z \to f \bar f)
\frac{\alpha}{\pi}\left[
-\frac{1}{2} \ln^2 \left(\frac{\mu^2}{s}\right)
-\frac{3}{2} \ln   \left(\frac{\mu^2}{s}\right)
+\frac{7}{18} \pi^2
+\frac{\pi}{2 \sqrt{3}}
-\frac{7}{2}
\right]
\,,
\end{equation}
where $\Gamma_0(Z \to f \bar f)$ is the tree level result.

In the computation of the real part, all final state particles have a common mass $\mu$ and the amplitude squared is integrated over the 3-body phase-space, parametrized as
\begin{equation}
\int d \Phi_3 = \frac{\pi^2}{4s} \int ds_{12} ds_{23}\,,
\end{equation}
where $\sqrt{s}$ is the center of mass energy of the decaying particle and 
$s_{ij}$ are two of the three possible final state two-body invariant masses.
It is convenient to introduce the variables
\begin{equation}
x= \frac{s_{12}-\mu^2}{s}\,,~~z= \frac{s_{23}-\mu^2}{s}\,,
\end{equation}
in terms of which the relevant bremsstrahlung integrals are
\begin{eqnarray}
\label{eq:reaint}
I_1 &=& \int_R dx dz \frac{1}{x^2} = \int_R dx dz \frac{1}{z^2}\,, \nonumber \\
I_2 &=& \int_R dx dz \frac{1}{xz}\,,  \nonumber \\
I_3 &=& \int_R dx dz \frac{1}{x} = \int_R dx dz \frac{1}{z}\,, \nonumber \\
I_4 &=& \int_R dx dz \frac{x}{z} = \int_R dx dz \frac{z}{x}\,, 
\end{eqnarray}
where R is the full available phase-space. They are reported in appendix~\ref{sec:appb}, and give 
\begin{equation}
\label{eq:realres}
\Gamma_{\rm R}(Z \to f \bar f)= \Gamma_0(Z \to f \bar f)
\frac{\alpha}{\pi}\left[
\frac{1}{2} \ln^2 \left(\frac{\mu^2}{s}\right)
+\frac{3}{2} \ln   \left(\frac{\mu^2}{s}\right)
-\frac{7}{18} \pi^2
-\frac{\pi}{2 \sqrt{3}}
+\frac{17}{4}
\right]
\,.
\end{equation}
By summing eqs.~\ref{eq:virtres} and \ref{eq:realres} one obtains
\begin{equation}
\label{eq:totres}
\Gamma(Z \to f \bar f)= \Gamma_0(Z \to f \bar f)\left(1+\frac{3}{4}
\frac{\alpha}{\pi} \right)\,,
\end{equation}
which is the expected result.

In eq.~\ref{eq:realres}, I used the following sums over fermion and photon polarizations
\bqa
\sum_{pol} u(p) \bar u(p) = \rlap/ p + \mu \,,~~~ 
\sum_{pol} v(p) \bar v(p) = \rlap/ p - \mu \,,~~~ 
\sum_{pol} \epsilon_\alpha(p) \epsilon^{\ast}_\beta(p) = 
- g_{\alpha\beta}\,.
\eqa
Due to the consistent FDR $\mu$ insertion, gauge invariance is kept; thus, any $p_\alpha p_\beta$ term added in the last equation does not contribute.
\section{Conclusions}
The FDR approach discriminates between observable physics and unobservable infinities occurring at large values of the integration momenta.
This interpretation allows one to define, in a mathematically consistent way and at any order in the perturbative expansion, the UV divergent integrals appearing in quantum field theories as four-dimensional integrals over the physical spectrum only. The physics of renormalizable theories is reproduced and the possibility for the non-renormalizable theories to become predictive is opened.
Infrared and collinear divergences can also be naturally accommodated. 

FDR looks promising for realistic calculations too. No counterterms need to be added to the Lagrangian, and both virtual and real contributions are kept in four dimensions, which may be particularly interesting at two (or more) loops, where subtracting powers of $1/\epsilon$ can become cumbersome in DR. These more practical aspects will be investigated in the near future.

\appendix
\section{1- and 2-point one-loop scalar integrals in FDR}
\label{sec:appa}
In this appendix, I explicitly compute the FDR scalar integrals
$I_0^{\rm FDR}$ and $I_2^{\rm FDR}$ of eqs.~\ref{eq:eq4} and ~\ref{eq:eq7}.  
By means of a common Feynman parametrization one rewrites
\begin{eqnarray}
K \equiv \int d^4q \left(\frac{1}{{\bar D}^2 \bar{q}^2} 
+ \frac{1}{{\bar D} \bar{q}^4} \right) =
\Gamma(3) \int_0^1 d \alpha
\int d^4q \frac{1}{(q^2-M^2\alpha-\mu^2)^3}\,,
\end{eqnarray}
where ${\bar D}= (\bar{q}^2-M^2)$ and $\bar{q}^2= q^2-\mu^2$.
Performing the integral in $d^4q$ gives
\begin{eqnarray}
K = -i \pi^2
\int_0^1 d \alpha
\frac{1}{M^2 \alpha +\mu^2} 
=  -\frac{i \pi^2}{M^2} \ln \frac{M^2+\mu^2}{\mu^2} \,,
\end{eqnarray}
from which eq.~\ref{eq:eq4} immediately follows.
Analogously,
\begin{eqnarray}
\int d^4q \frac{1}{{\bar D} \bar{q}^4} &=& \Gamma(3) \int_0^1 d \alpha
(1-\alpha)  \int d^4q \frac{1}{(q^2-M^2 \alpha-\mu^2)^3}
\nonumber \\
&=& -i \pi^2
\int_0^1 d \alpha
\frac{(1-\alpha)}{M^2 \alpha +\mu^2} 
= \frac{i \pi^2}{M^4}\left( 
M^2-(M^2+\mu^2) \ln \frac{M^2+\mu^2}{\mu^2}
\right)\,,
\end{eqnarray}
which gives eq.~\ref{eq:eq7}.
The same results can be obtained via eq.~\ref{eq:fdrconn}, by expressing 
$I_0^{\rm FDR}$ and $I_2^{\rm FDR}$ as the difference of two DR integrals.

\section{Virtual and real integrals for $\Gamma(Z \to f \bar f)$}
\label{sec:appb}
The one-loop scalar integrals appearing in the computation of $\Gamma(Z \to f \bar f)$ are (see eq.~\ref{eq:virtint})
\bqa
\label{eq:appb1}
B(s)&=&  i \pi^2 \left[ \ln\left(-\frac{\mu^2-i\epsilon}{s}\right) +2  \right]\,, \nonumber \\
B_0 &=&  -i \pi^2 \left( \frac{\pi}{\sqrt{3}}-2 \right)\,, \nonumber \\
B_1 &=&  -\frac{1}{2}B_0\,,  \nonumber \\
B^\prime_0 &=& \frac{i \pi^2}{\mu^2}
\left( \frac{2}{3}\frac{\pi}{\sqrt{3}}-1 \right)\,,  \nonumber \\
B^\prime_1 &=& -\frac{1}{2} B^\prime_0\,,  \nonumber \\
C(s)        &=&   \frac{i \pi^2}{s}
\left[\frac{1}{2}\ln^2\left(-\frac{\mu^2-i\epsilon}{s}
\right)
+ \frac{\pi^2}{9} \right]\,,\nonumber \\
C_{\rm R}  &=&  \frac{i \pi^2}{2}\,. 
\eqa
As for the real part, the integrals in eq.~\ref{eq:reaint} read
\begin{eqnarray}
\label{eq:appb2}
I_1 &=& \frac{s}{\mu^2}\left(\frac{2}{3} \frac{\pi}{\sqrt{3}} -1 \right)\,,\nonumber \\
I_2 &=& \frac{1}{2} \ln^2 \left(\frac{\mu^2}{s}\right) -\frac{7}{18}\pi^2\,, \nonumber \\
I_3 &=& - \ln \left(\frac{\mu^2}{s}\right) -1 -\frac{\pi}{\sqrt{3}}\,, \nonumber \\
I_4 &=& -\frac{3}{4}-\frac{1}{2} \ln \left(\frac{\mu^2}{s}\right)-\frac{\pi}{2 \sqrt{3}}\,.
\end{eqnarray}
All terms which vanish in the limit $\mu \to 0$ are neglected
in eqs.~\ref{eq:appb1} and \ref{eq:appb2}.

\acknowledgments
I thank G. Passarino for his comments on the manuscript and York Schroeder for pointing out redundancies in a previous version of figure~\ref{fig:fig4}. 
This work was performed in the framework of the ERC grant 291377, ``LHCtheory: Theoretical predictions and analyses of LHC physics: advancing the precision frontier''. I also thank the support of the MICINN project FPA2011-22398 (LHC@NLO).

\bibliography{fdr}{}
\bibliographystyle{JHEP}
\end{document}